\documentclass[prb,preprintnumbers,amsmath,amssymb,twocolumn]{revtex4}

\usepackage{graphicx}
\usepackage{dcolumn}
\usepackage{bm}
\usepackage{latexsym,epsfig}
\usepackage{chemarrow}

\begin{document}

\title{Unconventional Superfluidity in a model of Fermi-Bose Mixtures}

\author{K Sheshadri$^{1,}$} \email{kshesh@gmail.com}  
\author{A Chainani$^{2,}$} \email{chainania@gmail.com}

\affiliation{ $^{1}$226, Bagalur, Bangalore North, Karnataka State, India 562149 }

\affiliation{ $^{2}$Condensed Matter Physics Group, National Synchrotron Radiation Research Center, Hsinchu 30076, Taiwan }

\date{\today}

\begin{abstract}

A finite-temperature ($T>0$) study of a model of a mixture of spin-zero hardcore bosons and spinless fermions, with filling fractions $\rho_B$ and $\rho_F$, respectively, on a two-dimensional square lattice with composite hopping $t$ is presented. The composite hopping swaps the locations of a fermion and a boson that occupy nearest-neighbor sites of the lattice. The superfluid order parameter $\psi$, the femion hopping amplitude $\phi$, the chemical potential $\mu$, the free energy minimum $\tilde{F}$ and entropy $S$ are calculated in the limit $\rho_B+\rho_F=1$ within a mean-field approximation, and lead to a phase diagram in the $\rho_F - T$ plane. This phase diagram consists of a metallic superfluid phase under a dome-shaped $T(\rho_F)$, and insulating normal liquid and insulating normal gas phases outside the dome. These phases are  separated by coupled discontinuous transitions as indicated by jumps in $\psi$ and $\phi$. The maximum critical transition temperature $T_c$ is observed very close to $\rho_F = 1/2$. While $\tilde{F} (T)$ is continuous with a derivative discontinuity at $T=T_c (\rho_F)$ for $0 <\rho_F \le 1/2$ (first-order transition), it becomes {\em discontinuous} for $\rho_F>1/2$ (zeroth-order transition), where the entropy becomes negative for a range of temperatures below $T_c$. The ratio of $T_c$ to Fermi band width agrees remarkably with the ratio of $T_c$/$T_F$ (where $T_F$ is the Fermi temperature) of unconventional superfluids and superconductors like Fermi-Bose mixtures, the high-$T_c$ cuprates, iron-based and hydride superconductors, that exhibit experimental values of $T_c$ spread over nine orders of magnitude from $\sim 200$nK to $\sim 260$K.

\end{abstract}



\maketitle

\section{Introduction}
\label{introduce}

Fermi-Bose mixtures (FBMs) constitute an unusual and important state of matter, including well known examples like He$^3$-He$^4$ mixtures \cite{Ebner}, the mixed phase of type-II superconductors, ultracold atom systems\cite{Truscott, Schreck, Hadzibabic}, unconventional superconductors which exhibit Bardeen-Cooper-Schrieffer to Bose-Einstein condensation (BCS-BEC) crossover\cite{Lubashevsky, Okazaki, Kasahara, Rinott}, and so on. 
Experimental and theoretical studies of FBMs have shown remarkable results, particularly in terms of the BCS-BEC crossover across a Feshbach resonance  \cite{Regal}, that have revealed their distinct aspects compared to the limiting cases  of BCS superconductivity and BEC superfluidity \cite{Randeria, Ketterle}. 

The BCS-BEC crossover was originally predicted to occur for excitons in semiconductors \cite{Keldysh} and quarks in high energy physics\cite{Kerbikov}. However, it was first reported experimentally in ultracold fermionic atoms with s-wave interactions \cite{Bartenstein}. Unusual and unexpected results  include formation of a Feshbach molecule\cite{Cumby}, and the role of three body physics\cite{Bloom} in FBMs.  On the other hand, the role of BCS-BEC crossover in condensed matter involves experimental results on iron-based superconductors\cite{Lubashevsky, Okazaki, Kasahara, Rinott} and its relation to well-accepted theoretical results\cite{Randeria, Randeria2, quick}. While interactions between fermions mediated by phonons define the BCS theory of superconductivity, several studies have also considered their importance in mixtures of ultracold atoms \cite{Bijlsma, Viverit, Capuzzi, Albus}. The Boson-Fermion (BF) model \cite{Schafroth}, which preceded the BCS theory, discusses itinerant fermions hybridizing with bosons composed of bound pairs of fermions of opposite spins. The BF model was subsequently used to study electrons interacting with local lattice deformations \cite{Ranninger} as well as high temperature superconductivity \cite{Ranninger2, Friedberg, Gershkenbein, Domanski}. Recent studies have applied it for describing resonance superfluids in the BCS-BEC crossover regime \cite{Shin}, as well as a temperature driven crossover in an FBM \cite{Maska2}. These studies have shown the importance and interplay of bosonic and fermionic degrees of freedom in various physical systems.

Early studies on mixtures investigated the role of an attractive interaction between fermions and bosons. It was shown that an FBM with attractive interactions undergoes a collapse when the fermion number exceeded a critical value\cite{Ufrecht}. The breakthrough in controlling a Feshbach resonance in FBMs allowed researchers to effectively tune the boson-fermion interaction and control the system from collapsing at high densities \cite{Ospelkaus}. In contrast, theoretical studies employing repulsive interactions could describe various stable density  configurations of FBMs. The role of finite temperatures and going beyond the mean-field approximation was also investigated \cite{Viverit}. In the case of a strongly repulsive quasi one-dimensional FBM,\cite{Guan} it was shown that the phase diagram as a function of applied magnetic field $H$ displays a pure boson phase for $H = 0$,  polarized fermions and bosons coexisting  for $0 < H <  H_c$, and a fully polarized fermion phase for $H > H_c$. More interestingly, for an FBM on a 2D optical lattice in the framework of an extended single band Hubbard model with Coulomb interaction terms between bosons ($U_{BB}$), between fermions ($U_{FF}$) and an additional Coulomb interaction between bosons and  fermions ($U_{BF}$), it was shown that the bosons can mediate an attractive interaction between fermions, leading to fermion paired states with different $s, ~p$ and $d$ orbital symmetries \cite{Wang}. Further, the phase diagram as a function of $U_{BF}$ versus fermion number also revealed the existence of spin density wave and charge density wave phases. The authors also predicted that for experimentally accessible regime of parameters, the 2D FBM would exhibit superfluidity with an unconventional fermion pairing having a transition temperature around a percent of the Fermi energy. On the other hand,  the role of interaction-dependent temperature effects in an FBM were investigated by Cramer\cite{Cramer}. It was shown that adiabatic temperature changes of the FBM occur which depend on the interaction between fermions and bosons\cite{Cramer}.

In addition, the dynamics of FBMs  has also been investigated and it was shown that long range density wave phases can be obtained for fermions and bosons hopping independently in the presence of on-site boson-boson $U_{BB}$ and boson-fermion $U_{BF}$ Coulomb interactions \cite{Lewenstein, Pollet}. However, a  {\em composite}  hopping that exchanges a fermion with a boson, when they occupy neighboring sites, was not considered in earlier work. This form of hopping was proposed recently by us \cite{zeroTarxiv, zeroTPRO} and distinguishes our work from earlier work on FBMs. 
In this work, we calculate the thermodynamic properties of a model of FBM on a two-dimensional square lattice with composite hopping between neighboring spinless fermions and hardcore bosons, extending our earlier study of $T=0$ properties \cite{zeroTarxiv, zeroTPRO}. As in the previous work, we use a mean-field approximation and restrict ourselves to the case 
\begin{equation}
\label{eq:fill1}
\rho_F + \rho_B = 1,
\end{equation}
where $\rho_F$ and $\rho_B$ are the filling fractions of the fermions and bosons, respectively. To recall, at $T=0$, the model displays two distinct phases separated by {\em coupled first-order} transitions at Fermi filling fraction $\rho_F \simeq 0.3$: for $\rho_F<0.3$ the Fermi sector is insulating and the Bose sector is a normal liquid, while for $\rho_F>0.3$ the Fermi sector is metallic and the Bose sector is a superfluid. In the present work, we find that thermal fluctuations suppress superfluidity, and at a certain $T=T_c (\rho_F)$ there is a discontinuous transition to an insulating non-superfluid phase as shown by the superfluid amplitude $\psi (T)$ and the fermion hopping amplitude $\phi (T)$. We further find that the transition occurring at $T_c (\rho_F)$ is first order for $0.3<\rho_F \le 1/2$ (the minimum free energy  $\tilde{F}(T)$ is continuous with a discontinuity in its first derivative), but is zeroth order for $1/2<\rho_F<1$ (the minimum free energy  $\tilde{F}(T)$ is discontinuous). In the latter regime, the entropy becomes {\em negative} for a range of temperatures below $T_c$. We compute the ratio of $T_c$ to the Fermi band width, and find remarkable agreement with measured values in the range of $0.02$ to $0.20$ for a wide variety of unconventional superfluids and superconductors, including Fermi-Bose mixtures, the high-$T_c$ cuprates, iron-based superconductors and hydrides, that have their $T_c$ spread over nine orders of magnitude from a few hundred nanokelvins to a few hundred kelvins\cite{zeroTPRO}. Our estimate for the superconducting $T_c$ in the solid-state context with known experimental band widths or the Fermi temperature are consistent with observed $T_c$'s of the cuprates as well as iron-based superconductors.

\section{The Composite-Hopping Model and Its Mean-Field Thermodynamics}
\label{calculate}

We consider the composite-hopping model with Hamiltonian
\begin{eqnarray}
\label{eq:ham}
H &=& -\alpha \sum_{i} \left[b_i^{\dagger}b_i + f_i^{\dagger}f_i - 1\right]  -\mu \sum_{i} f_i^{\dagger}f_i   \nonumber \\
&& - t \sum_{<ij>} f_i^{\dagger}f_j b_j^{\dagger} b_i
\end{eqnarray}
that was proposed in a recent study for $T=0$\cite{zeroTarxiv, zeroTPRO}, where the notation used above is also explained. In this work also, we consider a FBM on a two-dimensional square lattice. The composite hopping term, the last term above, results in swapping of a hardcore boson and a spinless fermion when they occupy nearest-neighbor sites. Using the mean-field approximation, this term is  transformed according to 
\begin{eqnarray}
\label{eq:mfa}
f_i^{\dagger}f_j b_j^{\dagger} b_i &\simeq& \langle f_i^{\dagger}f_j \rangle ~(\langle b_j^{\dagger} \rangle b_i + \langle b_i \rangle b_j^{\dagger} - \langle b_j^{\dagger} \rangle \langle b_i \rangle)  \nonumber \\
&& + \langle b_j^{\dagger} \rangle \langle b_i \rangle f_i^{\dagger}f_j - \langle f_i^{\dagger}f_j \rangle \langle b_j^{\dagger} \rangle \langle b_i \rangle,
\end{eqnarray}
so $H$ is approximated by a mean-field Hamiltonian
\begin{eqnarray}
\label{eq:mfham}
H^{MF} &=& H_0 + H_1 + H_2,  ~~\mathrm{where} \nonumber \\
H_0 &=& N (2 \phi \psi^2 + \alpha),  \nonumber \\
H_1 &=& - (\alpha+\mu)\sum_{i} f_i^{\dagger}f_i - \frac{1}{z}\psi^2 \sum_{<ij>} f_i^{\dagger}f_j, ~ \mathrm{and}  \nonumber \\
H_2 &=& \sum_{i} \left[ -\alpha b_i^{\dagger}b_i - \phi \psi (b_i+b_i^{\dagger}) \right].
\end{eqnarray}
We have taken $zt=1$ ($z$ is the coordination number of the lattice), and introduced the thermodynamic expectation values
\begin{equation}
\label{eq:sce}
\phi = \langle f_i^{\dagger} f_j \rangle, ~~ \psi = \langle b_i \rangle = \langle b_j^{\dagger} \rangle.
\end{equation}
We assume $\phi ~\mathrm{and} ~\psi$ to be real and homogeneous and consider $\psi$ to be the superfluid order parameter \cite{FisherWeichman89, shesh93, shesh95, gutz, spin1_1, spin1_2, spin1_3}. For the hardcore bosons, we use the single-site boson occupation number basis $\{ |0\rangle, ~ |1\rangle \}$ for diagonalizing the $2 \times 2$ matrix  $h_2$ of $H_2/N$, i.e.,
\begin{equation}
\label{eq:matrix}
h_2 = 
\begin{bmatrix}
    0       & -\phi \psi  \\
    -\phi \psi       & -\alpha 
\end{bmatrix}
\end{equation}
that has the eigenvalues
\begin{equation}
\label{eq:bspect}
\lambda_{\pm} = \frac{1}{2} \left[ -\alpha \pm R \right], ~~\mathrm{where} ~~R  =\sqrt{\alpha^2 + 4\phi^2\psi^2}.
\end{equation}
Using the Fourier transform
\begin{equation}
\label{eq:ft}
f_i = \frac{1}{\sqrt{N}} \sum_{\bf k} e^{i {\bf k.r_i}} f_{\bf k},
\end{equation}
the Hamiltonian $H_1$ of the fermion sector becomes
\begin{equation}
\label{eq:h1ft}
H_1 = \sum_{\bf k} (\varepsilon_{\bf k}-\mu) f_{\bf k}^{\dagger} f_{\bf k},
\end{equation}
where
\begin{eqnarray}
\label{eq:dispersion}
\varepsilon_{\bf k} &=& -\alpha - \psi^2 \gamma_{\bf k}, ~~\mathrm{and} \nonumber \\
\gamma_{\bf k} &=& \frac{2}{z} (\cos k_x + \cos k_y).
\end{eqnarray}

The free energy per lattice site $F=F(\alpha, \psi)$ is now
\begin{eqnarray}
\label{eq:fe}
F &=& \frac{1}{2} (\alpha - R) + 2 \phi \psi^2 -   T \ln (1+e^{-R/  T}) \nonumber \\
&& -   T \frac{1}{N} \sum_k \ln \left[ 1+e^{(\mu-\varepsilon_{\bf k})/  T} \right].
\end{eqnarray}
We take $k_B=1$ here and in the following. To calculate $\phi = \phi(\alpha, \psi)$, we use its definition in (\ref{eq:sce}) and go over to $k$-space to get
\begin{equation}
\label{eq:phi1a}
\phi = \frac{1}{Nz} \sum_{<ij>} \langle f_i^{\dagger} f_j \rangle = \frac{1}{N} \sum_{\bf k} \gamma_{\bf k} \langle f_{\bf k}^{\dagger} f_{\bf k} \rangle.
\end{equation}
By definition,
\begin{equation}
\label{eq:fdfunc}
\langle f_{\bf k}^{\dagger} f_{\bf k} \rangle = \frac{\mathrm{Tr} ( f_{\bf k}^{\dagger} f_{\bf k} e^{-H_1/T})}{\mathrm{Tr}( e^{-H_1/T})} = \frac{1}{1+e^{(\varepsilon_{\bf k}-\mu)/ T}},
\end{equation}
and so
\begin{equation}
\label{eq:phi1}
\phi = \frac{1}{N} \sum_{\bf k} \frac{\gamma_{\bf k}}{1+e^{(\varepsilon_{\bf k}-\mu)/ T}}.
\end{equation}
Using $\gamma_{\bf k} = -(\alpha + \varepsilon_{\bf k})/\psi^2$ (the first equation in (\ref{eq:dispersion})) we obtain
\begin{equation}
\label{eq:phi2}
\phi = -\frac{1}{N} \sum_{\bf k} \frac{\alpha + \varepsilon_{\bf k}}{\psi^2} \frac{1}{1+e^{(\varepsilon_{\bf k}-\mu)/ T}}.
\end{equation}
Introducing the density of states
\begin{equation}
\label{eq:dos}
\rho (E) = \frac{1}{N} \sum_{\bf k} \delta(E-\varepsilon_{\bf k}),
\end{equation}
we can write
\begin{equation}
\label{eq:phi3}
\phi = -\frac{1}{\psi^2}\int_{E_0}^{\mu} dE ~\frac{\alpha + E}{1+e^{(E-\mu)/ T}} \rho(E),
\end{equation}
where $\mu$ is chosen such that the fermion filling fraction
\begin{equation}
\label{eq:rhof}
\rho_F(\alpha, \psi) = \int_{E_0}^{\mu} dE \frac{1}{1+e^{(E-\mu)/ T}}\rho(E)
\end{equation}
has a desired value. Here, $E_0 = -\alpha-\psi^2$ is the minimum value of fermion energy. To calculate the density of states (\ref{eq:dos}),  we convert the $k$-sum in to an integral according to ~$(1/N)\sum_{\bf k} \to (1/4\pi^2) \int d{\bf k}$. Since $\varepsilon_{\bf -k} = \varepsilon_{\bf k}$, the $k$-space integral is four times the integral over the first quadrant of the Brillouin zone, and so we have
\begin{equation}
\label{eq:dos2}
\rho (E) = \frac{1}{\pi^2} \int_0^{\pi} dk_x \int_0^{\pi} dk_y ~\delta (E+\alpha+\psi^2\gamma_{\bf k}).
\end{equation}
The integral over $k_y$ can be easily evaluated, and we get
\begin{eqnarray}
\label{eq:dos3}
\rho (E) &=& \frac{2}{\pi^2\psi^2} f\left( \frac{\alpha + E}{\psi^2} \right), ~\mathrm{where} \nonumber \\
f(u) &=& \int_0^{\pi} \frac{dk_x}{\sqrt{1-(2 u + \cos k_x)^2}}.
\end{eqnarray}
We can readily see that the function $f(u)$ is real only when $-1 \le u \le 1$, and is non-negative. Therefore we have the inequality $-\alpha-\psi^2 \le E \le -\alpha+\psi^2$ for the fermion energy $E$. We substitute the above expression for $\rho (E)$ in to equations (\ref{eq:phi3}) and (\ref{eq:rhof}) and transform the integrals to obtain
\begin{eqnarray}
\label{eq:rhofphi}
\rho_F &=& \frac{2}{\pi^2}\int_{-1}^{u_F}  du \frac{f(u)}{1+e^{(u-u_F)\psi^2/ T}} ~\mathrm{and} \nonumber \\ 
\phi &=& -\frac{2}{\pi^2}\int_{-1}^{u_F}  du \frac{u f(u)}{1+e^{(u-u_F)\psi^2/ T}},
\end{eqnarray}
where $u_F=(\alpha+\mu)/\psi^2$. This helps us choose the value of $\mu$ for a desired $\rho_F$, given the values of $(\alpha, \psi)$. For any $(\rho_F, T)$, we determine $\psi$ and $\alpha$ by solving $\partial F/\partial \psi = 0 ~\mathrm{and} ~ \partial F/\partial \alpha = 0$ simultaneously. These two equations, and the two equations in (\ref{eq:rhofphi}) above, are solved iteratively to obtain $(\alpha, \psi)$ as well as $(\phi, \mu)$ for any chosen $(\rho_F, T)$. In general, there could be multiple solutions $(\alpha, \psi)$. We substitute each solution in $F(\alpha, \psi)$ and denote the resulting free energy minimum for the solution by $\tilde{F}$; the correct solution is the one that corresponds to the lowest $\tilde{F}$. From equation (\ref{eq:fe}), we obtain
\begin{eqnarray}
\label{eq:federivs}
\frac{\partial F}{\partial \psi} &=& 2 \psi (\phi + \phi_{\psi} \psi) \left[ 1-\frac{\phi}{R} \chi \right], ~ \mathrm{and}  \nonumber \\
\frac{\partial F}{\partial \alpha} &=& \frac{1}{2} \left[ (1-2\rho_F) - \frac{\chi}{R} \alpha \right] + 2 \phi_{\alpha} \psi^2 \left[ 1-\frac{\chi}{R} \phi \right], \nonumber \\
\end{eqnarray}
where $\phi_{\psi} = \partial \phi/\partial \psi, ~ \phi_{\alpha} = \partial \phi/\partial \alpha$, and
\begin{equation}
\label{eq:chi}
\chi(R,T) = \frac{e^{R/T}-1}{e^{R/T}+1}.
\end{equation}
Since $\phi ~\mathrm{and} ~ \phi_{\psi}$ are positive (see equations (\ref{eq:phi1}) and (\ref{eq:rhofphi})), we always have $\phi + \phi_{\psi} \psi>0$, so $\partial F/\partial \psi = 0$ gives
\begin{equation}
\label{eq:finiteT}
\psi = 0 ~~ \mathrm{or} ~~ R = \phi \chi(R,T).
\end{equation}
Using this in equation (\ref{eq:federivs}), $\partial F/\partial \alpha = 0$ gives
\begin{equation}
\label{eq:alpha}
\alpha = (1-2\rho_F) R/\chi.
\end{equation}

\section{Tractable Limits and Some General Observations}
\label{section:analyze}

Now that we have derived the implicit equations for $\psi$ and $\alpha$, we first analyze two simple limits: the $\psi=0$ (disordered phase) and the $T=0$ limits. We obtain closed-form solutions for $\alpha, ~\psi$ and $\tilde{F}$ in these limits. For the general case, we make some observations before moving on to a discussion of the numerical results in the next section. 

From equation (\ref{eq:alpha}), we get $\alpha^2 [\chi^2 - (1-2\rho_F)^2] = 4\phi^2\psi^2(1-2\rho_F)^2$. The solution $\psi=0$ corresponds to the disordered state, i.e., the Bose sector is in a non-superfluid, normal phase. We then get $\alpha=0$ or $\chi=|1-2\rho_F|$. In this state, $R=|\alpha|$, so using the definition of $\chi$ in equation (\ref{eq:chi}) we get
\begin{equation}
\label{eq:alphadis}
\alpha = 0, ~ T[\ln(1-\rho_F)-\ln\rho_F]
\end{equation}
in the disordered state ($\psi=0$). Of these two solutions for $\alpha$, we must pick the one corresponding to the lower $\tilde{F}$. Substituting the above in equation (\ref{eq:fe}) for $F$, we obtain $\tilde{F}_I = -T(2\ln2)$ for the first solution $\alpha=0$, and $\tilde{F}_{II} = -T[\ln2-\ln(1-\rho_F)]$ for the second solution $\alpha=T[\ln(1-\rho_F)-\ln\rho_F]$. We can see that $\tilde{F}_I<\tilde{F}_{II}$ for $\rho_F<1/2$ and $\tilde{F}_{II}<\tilde{F}_I$ for $\rho_F>1/2$. This shows that in the disordered phase, $\alpha = 0$ for $\rho_F \le 1/2$ and $\alpha=T[\ln(1-\rho_F)-\ln\rho_F]$ (which is the same as $\chi=|1-2\rho_F|$) for $\rho_F>1/2$. In Fig. (\ref{fig:fig1}) we show the behavior of $-\tilde{F}_D/T$ (which is the disordered-phase entropy $S_D$) as a function of $\rho_F$.

At $T=0$, we get $\tilde{F}=0$ for the disordered state (since both $\tilde{F}_I$ and $\tilde{F}_{II}$ vanish). For the ordered state, $\chi = 1$ so that
\begin{equation}
\label{eq:zeroT}
\psi_0^2 = \rho_F (1-\rho_F), ~~ \alpha_0 = (1-2\rho_F)\phi_0,
\end{equation}
and the minimum free energy is $\tilde{F}_0 =\rho_F^2[-\phi_0 + u_F(\rho_F-1)]$ by taking the $T \rightarrow 0$ limit in equation (\ref{eq:fe}) and substituting the above expressions for $\psi_0, ~ \alpha_0$. Here $\psi_0, ~\phi_0, ~\alpha_0$ and $\tilde{F}_0$ are $T=0$ values. This gives coupled first-order transitions at $\rho_F \simeq 0.3$ at $T=0$ between a metallic superfluid for $\rho_F>0.3$ and an insulating normal liquid for $\rho_F \le 0.3$. We referred to the latter as insulating normal gas in our earlier work\cite{zeroTarxiv, zeroTPRO}. However, since the destruction of superfluidity of the Bose sector in this regime is due to an interplay between correlation and quantum effects and not due to temperature, this unusual phase should be more appropriately called an insulating normal liquid rather than an insulating normal gas.

As we increase the temperature at a fixed $\rho_F$, thermal fluctuations suppress superfluidity, reducing $\psi$. To see how this happens, we rewrite the ordered-phase self consistency equation $R=\phi\chi$ in the form $T=J(\psi)$ where $J(\psi)=\phi x/\ln[(1+x)/(1-x)]$; here $x=\sqrt{(1-2\rho_F)^2+4\psi^2}$. The function $J(\psi)$ has zeros at $\psi=0$ (when $\phi=0$) and $\psi=\psi_0$ (when $x=1$). Since it is positive, it must have a maximum in the interval $(0, \psi_0)$. This is graphically illustrated in Fig. (\ref{fig:fig2}). As $T$ increases from zero, the line $y=T$ intersects the curve $y=J(\psi)$ at two points $\psi=\psi_1, \psi_2$. Since $J(\psi)$ is a single-valued function, we have $\psi_2 < \psi_1 < \psi_0$ and further, $\psi_1$ decreases and $\psi_2$ increases as $T$ is increased. The solution $\psi = \psi_1$ corresponds to the ordered minimum of the free energy. As the temperature increases further, $y=T$ increases, while the maximum of $J(\psi)$ decreases, and at a certain temperature $T_2 (\rho_F)$, the line $y=T$ becomes a tangent to $y=J(\psi)$ at the maximum, when $\psi_1=\psi_2$. For $T>T_2 (\rho_F)$, the self consistency equation $T=J(\psi)$ has no solution, and therefore $T_c \le T_2$.

For $T \le T_2$, when the ordered solution exists, there are two possibilities. (1) The ordered free energy minimum ($\tilde{F}_O$) remains lower than the disordered free energy minimum ($\tilde{F}_D$) for all $T \le T_2$. In this case the transition temperature is $T_c = T_2$, and occurs when the ordered minimum ceases to exist. The free energy minimum value is discontinuous at the transition, since the system switches to the only solution that exists, namely $\psi=0$. This corresponds to a {\em zeroth-order} transition. 
A zeroth-order phase transition has previously been considered in the theory of superfluidity and superconductivity \cite{Maslov2004}.
More recently, the reentrant phase
transition in black holes has been discussed as a zeroth-order transition.\cite{Gunasekaran, Altamirano01, Zou, Hennigar, Altamirano02, Amin}.

Our numerical computations obtain this scenario for $\rho_F>1/2$. (2) There exists a certain temperature $T_1 < T_2$ such that $\tilde{F}_O < \tilde{F}_D$ for $T < T_1$, and $\tilde{F}_O > \tilde{F}_D$ for $T > T_1$, with the two phases coexisting at $T_c = T_1$, which is a point of {\em first-order} transition. Our numerical results obtain this result for $\rho_F \le 1/2$.

We can show that the transition is indeed zeroth order for $\rho_F > 1/2$ based on the disordered-phase behavior of $\alpha$ that we derived above. Formally, the solution of the self-consistency equation $R=\phi\chi$ is $\psi^2 =(1/4) [\chi^2 - (1-2\rho_F)^2]$. As we saw above, $\chi=|1-2\rho_F|$ in the disordered phase when $\rho_F>1/2$, so the ordered solution $\psi>0$ does not exist in the disordred phase. If $T_c < T_2$, then we would have the ordered solution existing in the disordered phase for $T_c < T < T_2$, which is a contradiction. So we must have $T_c = T_2$ in this case, resulting in a zeroth-order transition.

For $\rho_F \le 1/2$, however, the ordered-phase self-consistency equation in the disordered phase (where $\alpha=0$) becomes $2\psi = (e^{2\phi\psi/T}-1)/(e^{2\phi\psi/T}+1)$, that has a finite $\psi$ solution for $T<T_2$. If $T_c = T_2$, then we would have an ordered solution existing in the disordered phase for $T>T_c=T_2$, a contradiction. We therefore have $T_c < T_2$ in this case. The transition criterion in this case is clearly $\tilde{F}_O=\tilde{F}_D$ since the ordered and disordered minima both exist at the transition. The transition for $\rho_F \le 1/2$ is therefore first order.

We now turn to the correlation function $C({\bf r}) = (1/N)\sum_i\langle \delta \rho_{{\bf r}_i} \delta \rho_{{\bf r}_i + {\bf r}} \rangle$ where $\delta \rho_{{\bf r}_i} = f_i^{\dagger}  f_i - \rho_F$. We obtain $C({\bf r}) = - n^2({\bf r})$ where $n({\bf r}) = (1/N) \sum_{\bf k} \langle  f_{\bf k}^{\dagger}  f_{\bf k}  \rangle  e^{i{\bf k}.{\bf r}}$, and $C_{\bf q} = -(1/N) \sum_{\bf k} \langle  f_{\bf k}^{\dagger}  f_{\bf k}  \rangle  \langle  f_{\bf k+q}^{\dagger}  f_{\bf k+q}  \rangle$ for the Fourier transform of $C({\bf r})$\cite{zeroTarxiv, zeroTPRO}. 
In our earlier work, we could show that $n({\bf r})$ shows a periodicity of twice the lattice spacing for $\rho_F=1/2$, while no apparent periodicity was obtained for other values of $\rho_F$. 
Using the expression for the Fermi function in Eq. (\ref{eq:fdfunc}), we can readily see that in the disordered phase ($\psi=0$), we obtain $C_{\bf q}=-1/4$. This shows that there is no density wave (DW) order in the insulating normal gas phase, and the DW order obtained for the $T=0$ metallic superfluid\cite{zeroTarxiv, zeroTPRO} is stabilized by the nesting of the Fermi surface and the superfluidity of the Bose sector.

\section{Discussion}
\label{discuss}

\begin{figure}

\includegraphics[width=0.45 \textwidth]{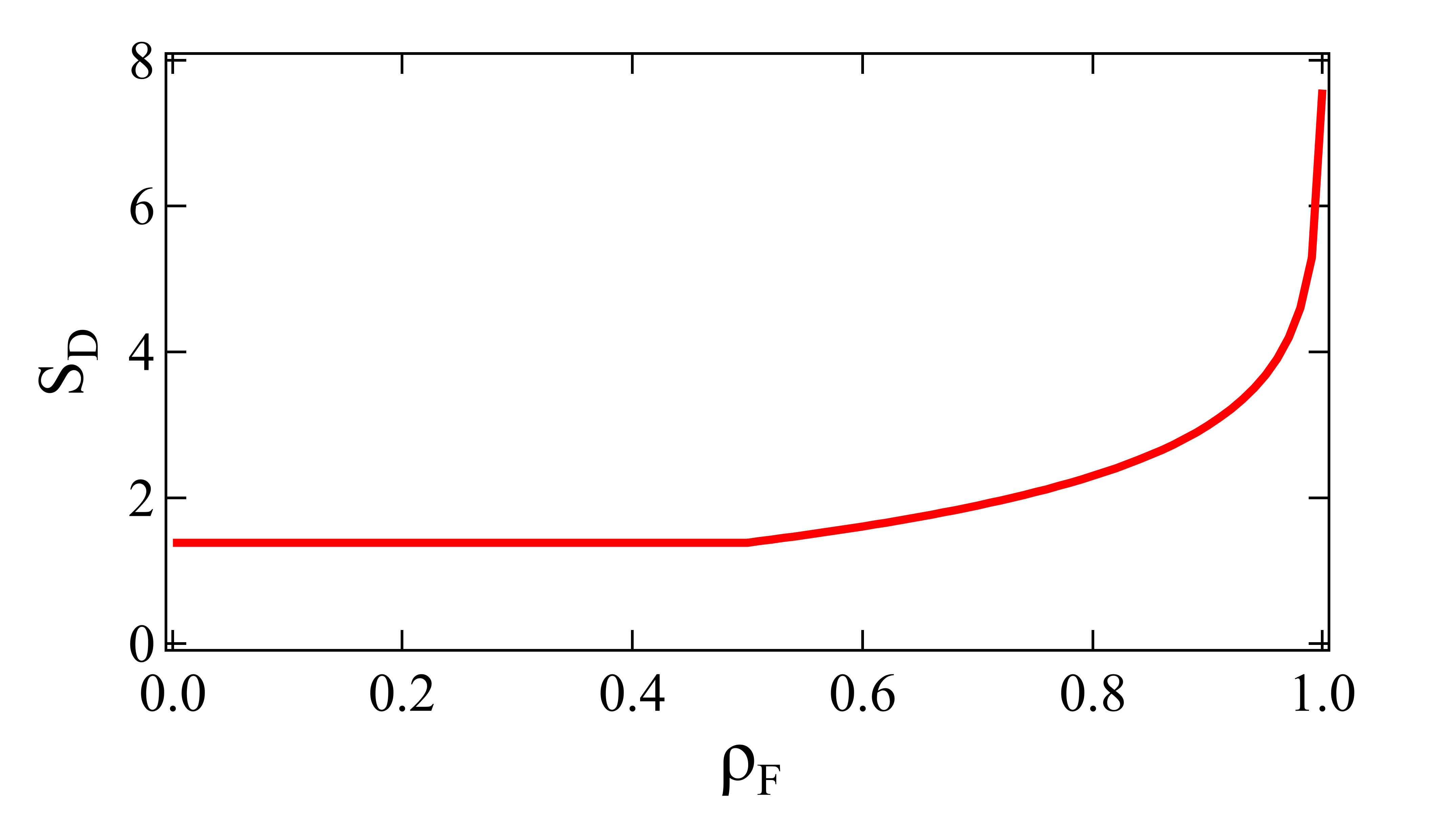}

\caption{The disordered-phase entropy $S_D = -\tilde{F}_D/T$ (equal to $-\tilde{F}_I/T$ for $\rho_F \le 1/2$ and $-\tilde{F}_{II}/T$ for $\rho_F>1/2$) plotted as a function of $\rho_F$. The high-temperature entropy is $2 \ln 2$, as expected, for $0 < \rho_F \le 1/2$. However, it is anomalously high for $1/2 < \rho_F \le 1$, the regime of zeroth-order transition.}
  	
 \label{fig:fig1}

\end{figure}

\begin{figure}

\includegraphics[width=0.45 \textwidth]{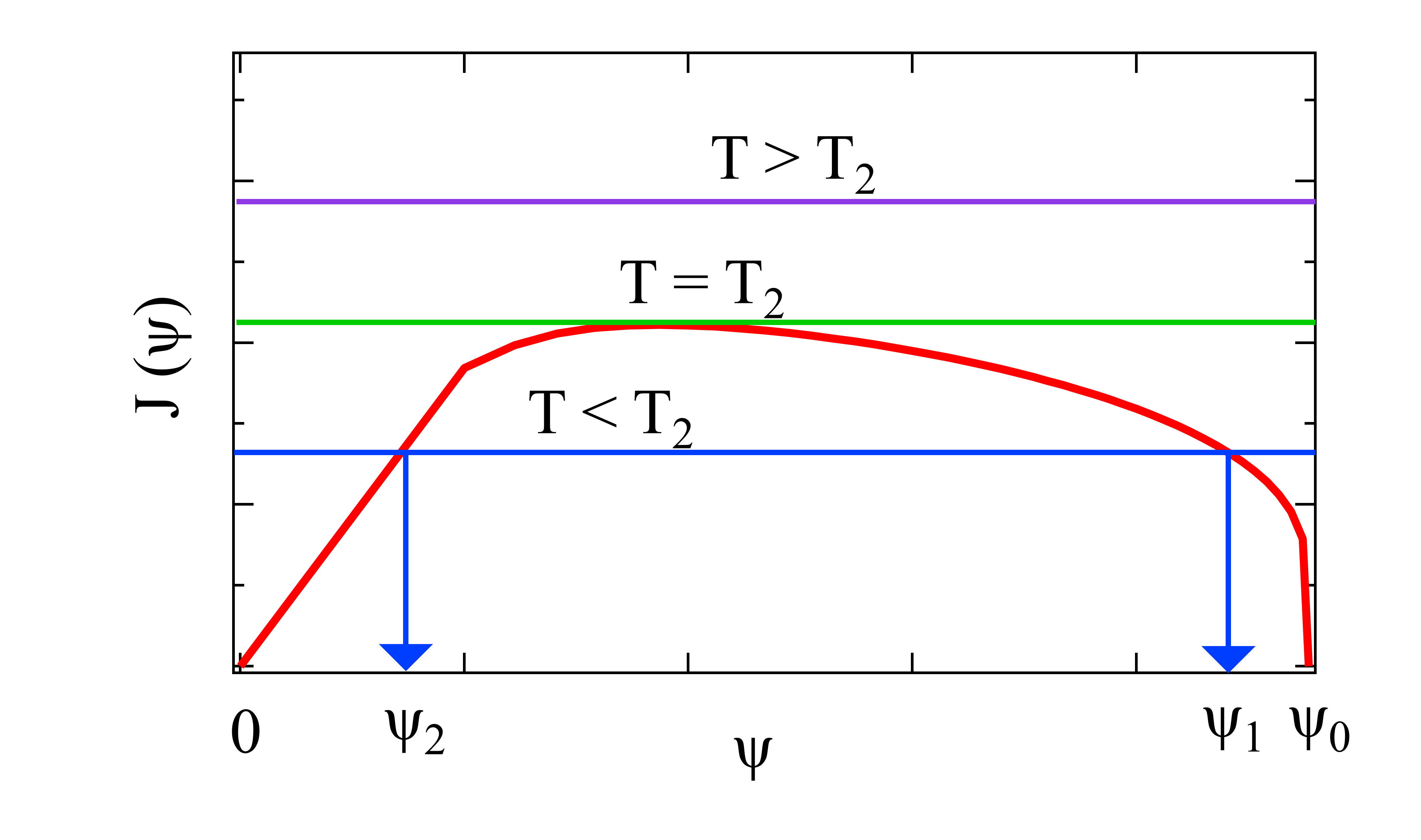}

\caption{A figure illustrating the graphical solution of the self-consistency equation $T=J(\psi)$. The function $J(\psi)$ vanishes at $\psi = 0, \psi_0$ and has a maximum in between. The horizontal lines show plots of $y=T$ (for $0<T<T_2, ~ T=T_2, ~ T>T_2$). For $0<T<T_2$, the line $y=T$ intersects the curve $y=J(\psi)$ (red curve) at two points $\psi=\psi_1, \psi_2$, with $\psi_0 \ge \psi_1 > \psi_2$, that correspond, respectively, to a minimum and a maximum of $F$. For $y=T_2$, the two points merge and the line is a tangent to the curve $J(\psi)$ at its maximum. For $T>T_2$, the self-consistency equation has no solution for any real value of $\psi$.}
  	
 \label{fig:fig2}

\end{figure}

\begin{figure*}

\centering \includegraphics[width=\textwidth]{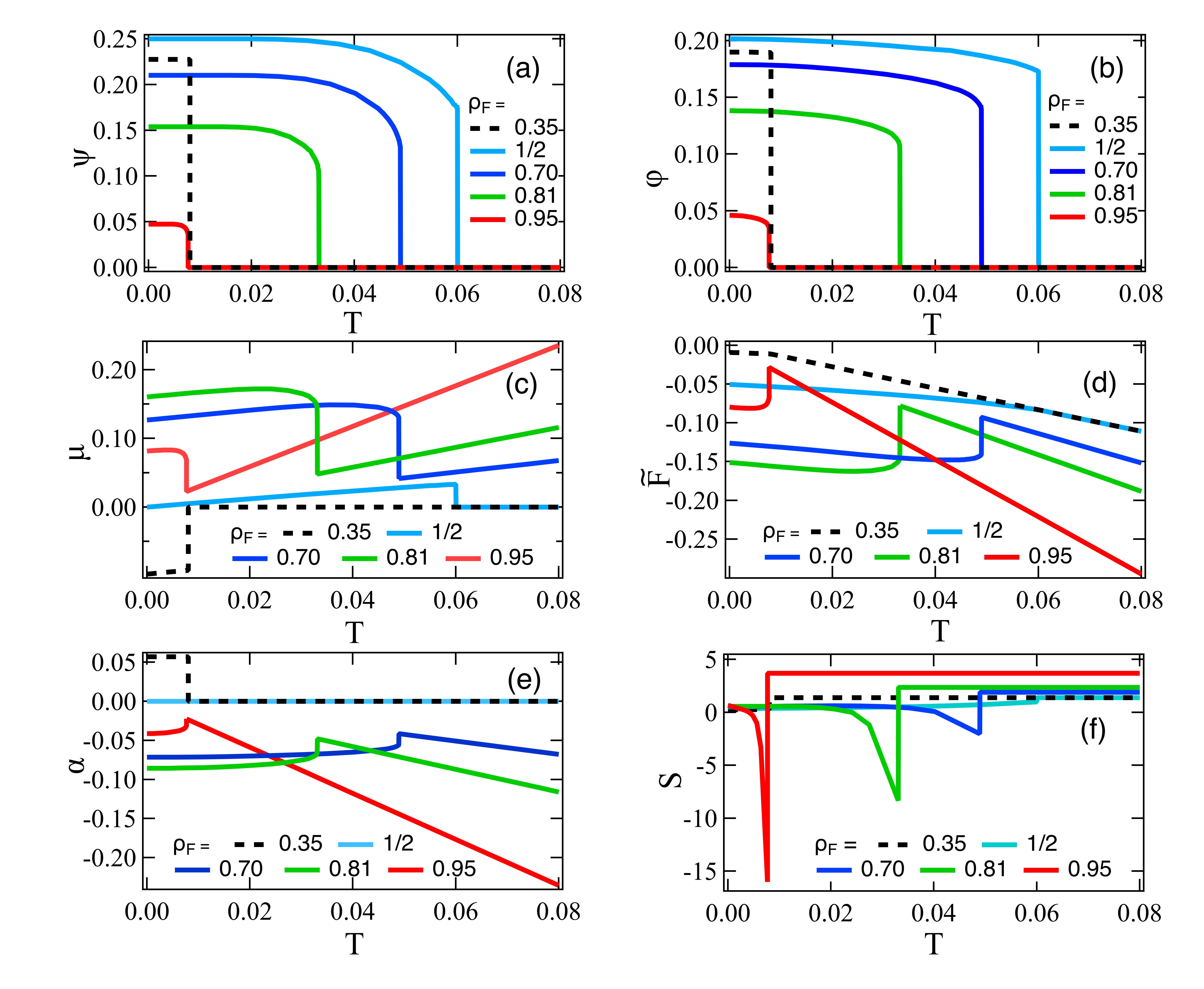}
\caption{Panels (a)-(f) show the $T$-dependence plots, respectively, of $\psi, ~\phi, ~ \mu, ~\tilde{F}, ~\alpha$ and $S$. Each panel has plots at five different values of $\rho_F$, namely $0.35, 1/2, 0.70, 0.81, 0.95$.}
  	
 \label{fig:fig3}

\end{figure*}

\begin{figure}

\includegraphics[width=0.45 \textwidth]{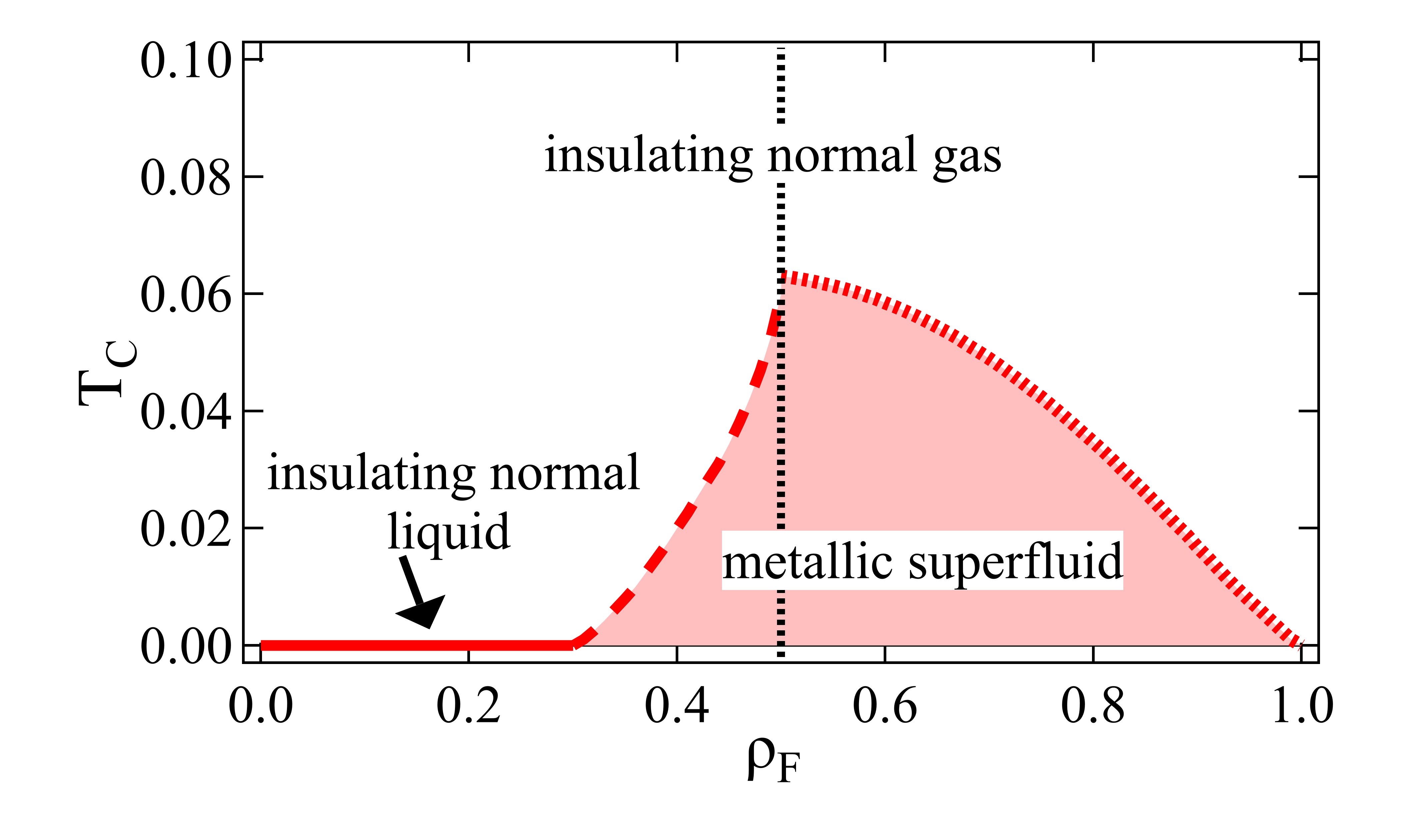}

\caption{The phase diagram of model (\ref{eq:ham}) in $(\rho_F, T)$ plane showing the metallic superfluid ($\phi, ~\psi>0$) and insulating non-superfluid phases ($\phi=\psi=0$) phases (the insulating normal liquid at $T=0$ and the insulating normal gas for $T>0$).  These phases are separated by lines of discontinuous transitions: at the phase boundary, the minimum free energy $\tilde{F}$ is continuous with a derivative discontinuity for $\rho \le 1/2$ (first-order transition, dashed line), while $\tilde{F}$ has a jump for $\rho>1/2$ (zeroth-order transition, dotted line).}
  	
 \label{fig:fig4}

\end{figure}

\begin{figure}

\includegraphics[width=0.45 \textwidth]{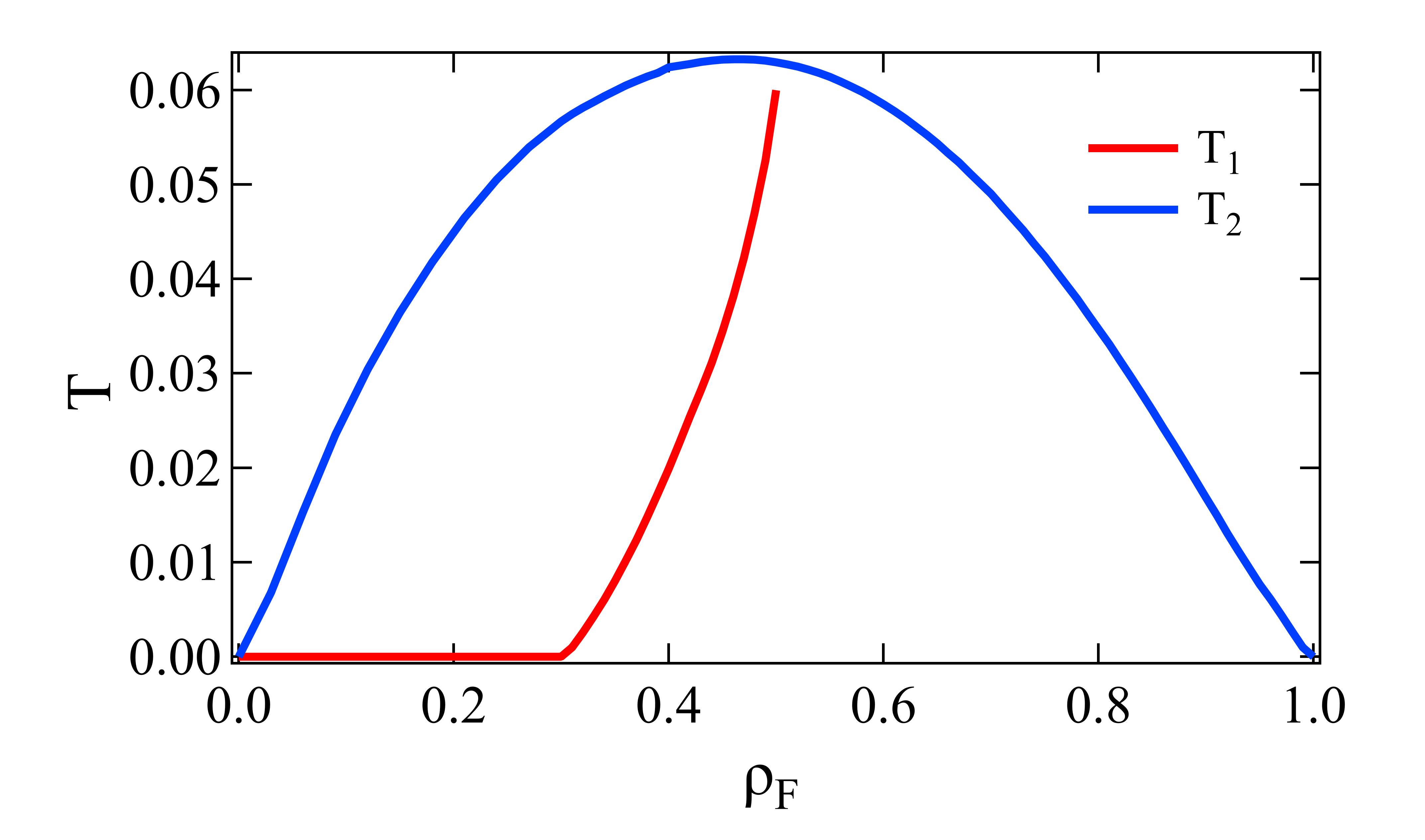}

\caption{The temperatures $T_1$ and $T_2$ are plotted as functions of $\rho_F$. As explained in the text, $T_1$ is the critical temperature for first-order transition, where ${\tilde F}_O={\tilde F}_D$. It is not defined for $\rho_F>1/2$, where ${\tilde F}_O < {\tilde F}_D$ for $T \le T_2$. The ordered minimum disappears at $T_2$. The critical temperature $T_c$ is $T_1$ for $\rho_F \le 1/2$ (first-order transition) and  $T_2$ for $\rho_F > 1/2$ (zeroth-order transition). Note that $T_2>T_1$. This is responsible for a segment of the phase boundary in Fig. \ref{fig:fig4} being vertical at $\rho_F=1/2$. }

\label{fig:fig5}

\end{figure}

\begin{figure}

\includegraphics[width=0.45 \textwidth]{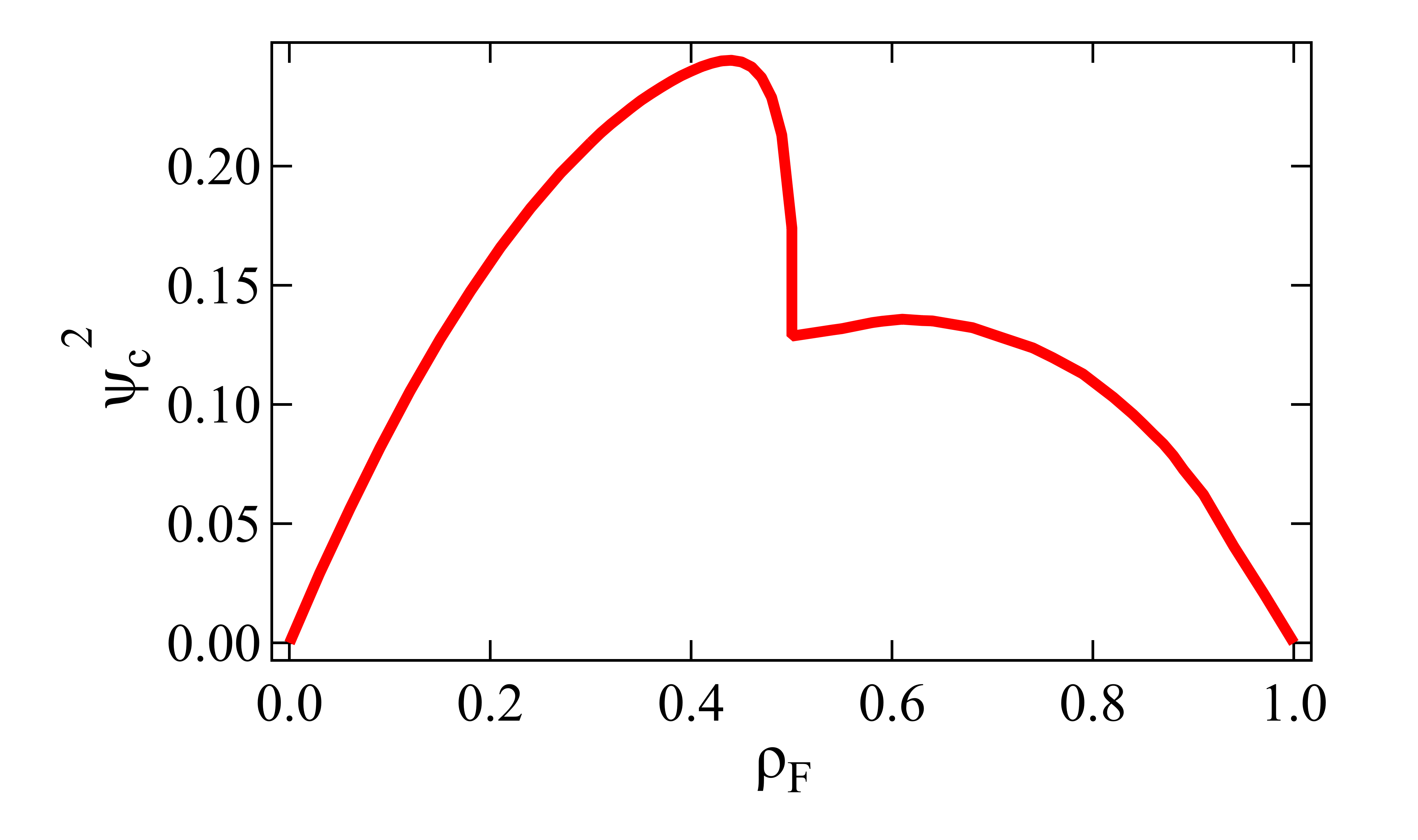}

\caption{ Plot of $\psi_c^2$, where $\psi_c$ is the jump in the superfluid order parameter at the discontinuous transition temperature $T_c$, as a function of $\rho_F$. }
 
\label{fig:fig6}

\end{figure}

\begin{figure}

\includegraphics[width=0.45 \textwidth]{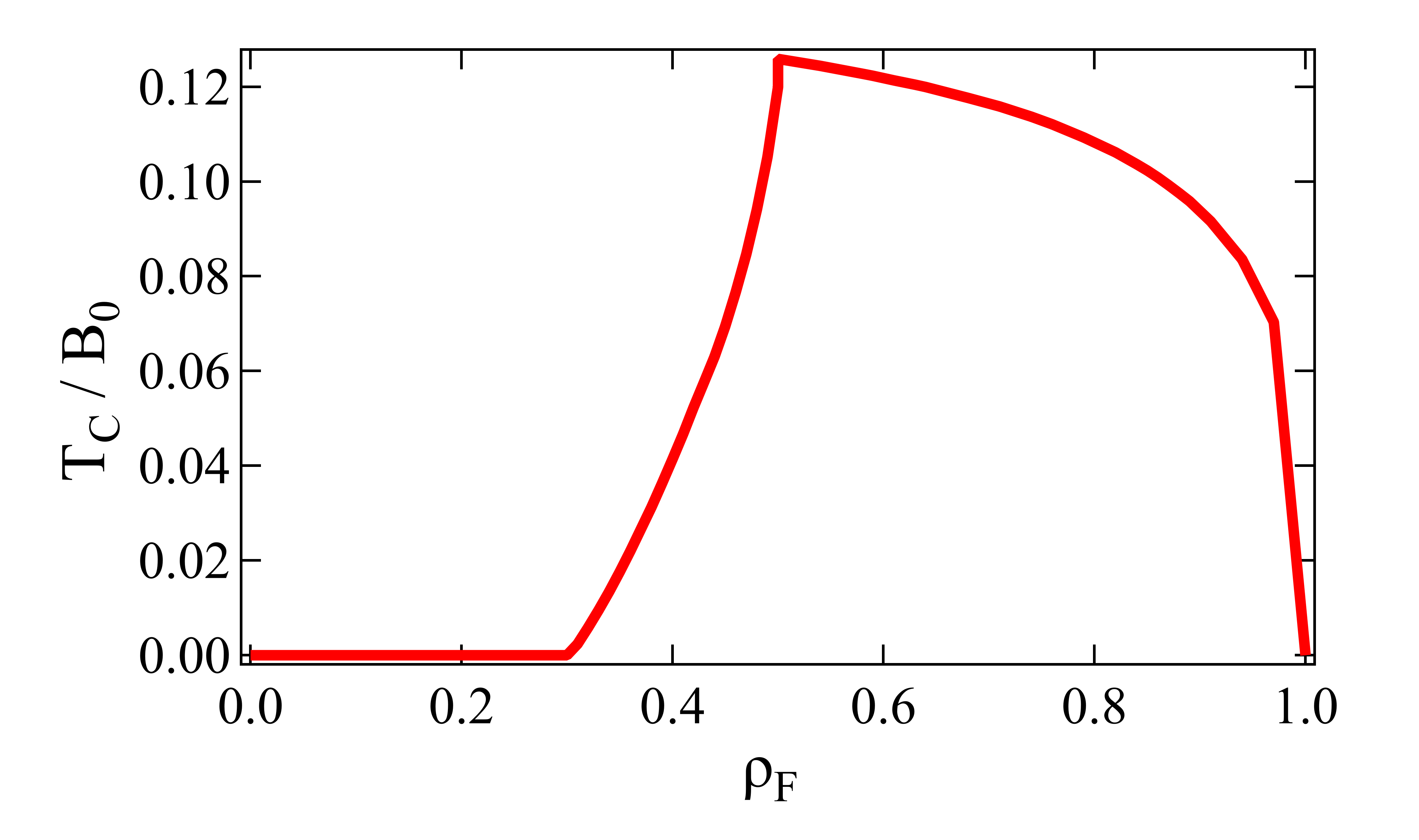}

\caption{ Plot of $T_c/B_0$ as a function of $\rho_F$. Here, $B_0 = 2\psi_0^2$ is the Fermi band width at $T=0$\cite{zeroTarxiv, zeroTPRO}.}
  	
\label{fig:fig7}

\end{figure}

\begin{figure}

\includegraphics[width=0.45 \textwidth]{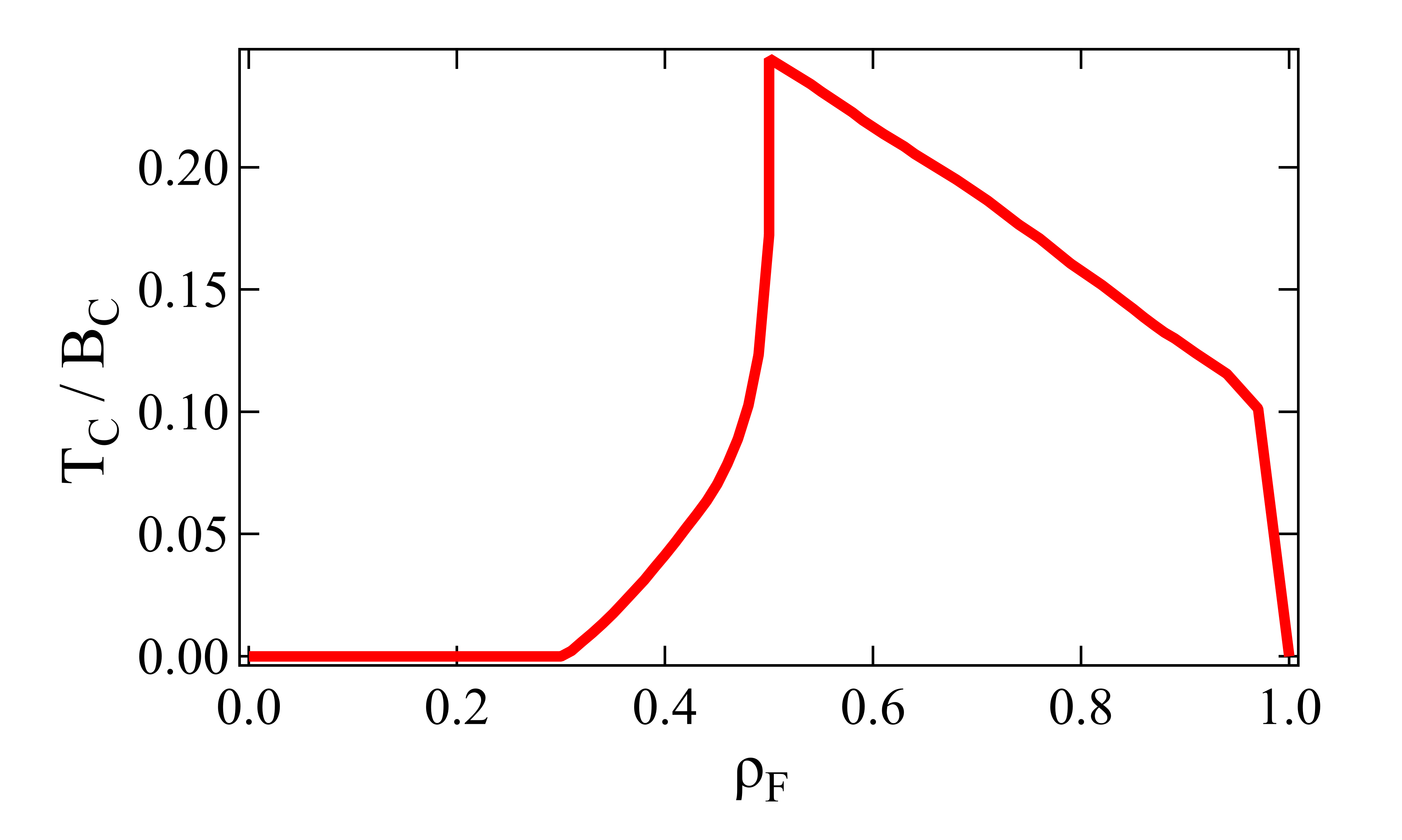}

\caption{ Plot of $T_c/B_c$ as a function of $\rho_F$. Here, $B_c = 2\psi_c^2$ is the $T=T_c$ analogue of the $T=0$ Fermi band widths $B_0$. }
  	
 \label{fig:fig8}

\end{figure}

Our numerical results are presented in figures 1-8. In Fig.\ref{fig:fig1} we show the behavior of disordered-phase entropy as a function of $\rho_F$. The entropy in this case is $S_D = -\tilde{F}_D/T$, where $\tilde{F}_D$ is the disordered-phase free energy minimum. As we discussed above, this is $-\tilde{F}_I/T$ for $\rho_F \le 1/2$ and $-\tilde{F}_{II}/T$ for $\rho_F>1/2$. We can see that the entropy is $2\ln 2$ for $\rho_F \le 1/2$ as one might expect in the disordered phase. However, the entropy is anomalously large for $\rho_F>1/2$, which is also the regime where the temperature-driven transition is zeroth order. We believe this is a consequence of the filling constraint (\ref{eq:fill1}), that leads to the solution in equation (\ref{eq:alpha}) for $\alpha$ so that $\alpha = T[\ln(1-\rho_F)-\ln\rho_F]$ in this case (see equation (\ref{eq:alphadis})).

To understand how temperature suppresses the metallic-superfluid order, we solve the self-consistency equations (\ref{eq:finiteT}) and (\ref{eq:alpha}) for $\psi, ~ \alpha$ using an approach graphically described in Fig.\ref{fig:fig2}. As shown in the figure, at a certain temperature $T<T_2$, the line $y=T$ intersects the curve $y=J(\psi)$ at two points $\psi_1, \psi_2 ~(\psi_0 \ge \psi_1 > \psi_2)$. The solution $\psi=\psi_1$ corresponds to the free energy minimum at this temperature. As the temperature increases, it is obvious from the figure that this solution moves to the left, i.e. decreases, leading to a thermal suppression of superfluidity. At a certain high temperature $T>T_2$, the line $y=T$ has no intersection with the curve $y=J(\psi)$. Therefore there is no solution to the self-consistency equation $T=J(\psi)$ at temperatures higher than a certain $T_2$, at which point the line $y=T$ becomes a tangent to the curve $y=J(\psi)$ at its maximum.

By numerically implementing this graphical method of solution, we can obtain $\psi, ~\phi, ~ \mu, ~\tilde{F}, ~\alpha$ and $S$ (the entropy) as the temperature $T$ is varied; these results are plotted in panels (a)-(f) of Fig.\ref{fig:fig3}. Each panel shows temperature dependence of one of these quantities for five different values of $\rho_F = 0.35, ~1/2, ~0.70, ~0.81, ~0.95$. This choice of $\rho_F$ values is the same as for the $T=0$ case, which span the superfluid metallic phase as reported earlier \cite{zeroTarxiv, zeroTPRO}. The figures show that at each of these fillings, there is a certain critical temperature $T_c$ where the model has a discrete phase transition: the quantities $\psi, \phi, \mu$ and $\alpha$ all show discontinuous changes at $T_c$ (figures \ref{fig:fig3}(a, b, c, e)). We can observe that at $\rho_F=0.35$ and $1/2$, the free energy minimum $~\tilde{F}$ is continuous, but with a derivative discontinuity, whereas it is discontinuous at $\rho_F=0.70, ~0.81$ and $0.95$ (Fig. \ref{fig:fig3}(d)).

In figure \ref{fig:fig3}(f) we show plots of entropy $S(T) = -\partial \tilde{F}/\partial T$, computed by numerical differentiation of $\tilde{F} (T)$. The unusual feature that can be readily seen is that the entropy becomes {\em negative} for certain temperatures below $T_c$ when $\rho_F>1/2$. 
While the concept of negative entropy has been applied to quantum information systems earlier \cite{cerf,delrio2011}, it's relevance for physical systems has been discussed only recently \cite{Chatzi2020}. Cerf and Adami showed that unlike in classical information theory, quantum conditional entropies can be negative for quantum entangled systems\cite{cerf}. Subsequently, del Rio et al. explained its thermodynamic meaning: negative entropy is related to a possible cooling of an environment connected to a quantum information system, when quantum information contained in the system is erased\cite{delrio2011}. In a very recent study\cite{Chatzi2020}, it has been proposed that the results of two independent inelastic neutron scattering experiments\cite{Olsen, Callear}, which showed an anomalous scattering from H$_2$ molecules in nanoscale confined geometries, can be explained on the basis of negative conditional entropy and quantum thermodynamics.  

Our numerical results for $\rho_F \le 1/2$ show that at $T=T_c=T_1$, the ordered and disordered phases coexist and we have $\tilde{F}_O = \tilde{F}_D$. The ordered minimum survives well into the disordered phase, and disappears at a temperature $T_2>T_1$. We have therefore a standard first-order transition in this case. On the other hand, for $\rho_F > 1/2$, the ordered minimum disappears abruptly at a certain temperature $T_2(\rho_F)$. The system then assumes the only minimum available to it, namely the disordered minimum. The free energy obviously shows a discontinuous jump in this case. The transition is therefore zeroth order. These numerical results are consistent with the qualitative remarks we made above in Section~\ref{section:analyze}.

As a result, we can define $T_1$ only for $\rho_F \le 1/2$, where clearly $T_c=T_1<T_2$. These results are summarized in figures \ref{fig:fig4} and \ref{fig:fig5}. Figure \ref{fig:fig4} shows the phase diagram of our model in $\rho_F - T$ plane, showing the metallic superfluid under the dome, and two insulating non-superfluid phases outside it. We refer to the insulating phase at $T=0$ for $0<\rho_F<0.3$ as an insulating normal liquid and the insulating phase for $T>0$ as an insulating normal gas. While the latter is dominated by thermal effects, the former is an unusual phase of bosons at $T=0$ that is a non-superfluid because of correlation and quantum effects.

Figure \ref{fig:fig5} shows the plots of $T_1 (\rho_F)$ and $T_2 (\rho_F)$. The function $T_2 (\rho_F)$ can be defined for $0 \le \rho_F \le 1$, and has a maximum at a certain $\rho_F < 1/2$. The temperature $T_1$ on the other hand remains lower than $T_2$; it vanishes for $0 \le \rho_F \le 0.3$, in agreement with the $T=0$ results\cite{zeroTarxiv, zeroTPRO}.

The line separating the two phases in Fig. \ref{fig:fig4} has a vertical segment at the van Hove point $\rho_F = 1/2$. This is as a consequence of two facts: (a) $T_c$ is the lower of $T_1$ and $T_2$, and (b) $T_1<T_2$, for $\rho_F \le 1/2$.

As we saw in Fig.\ref{fig:fig3}, the superfluid order parameter $\psi$ is discontinuous with a jump $\psi_c$ at $T_c$. In Fig.\ref{fig:fig6} we show the plot of $\psi_c^2 (\rho_F)$. The jump in the order parameter seems to be maximum around the same $\rho_F$ where $T_2 (\rho_F)$ is maximum. It shows an abrupt drop at $\rho_F=1/2$, and after a broad maximum, decreases slowly towards zero at $\rho_F=1$.

Based on the zero-temperature Fermi band width of $B_0 = 2\psi_0^2$, the ratio $T_c/B_0$ agreed very well with measured values for several types of unconventional superfluids and superconductors in our earlier work\cite{zeroTarxiv, zeroTPRO}. This value of $T_c \simeq 0.12$ in our earlier work was not based on a calculation, but only an estimate based on the zero-temperature free energy minimum for $\rho_F=1/2$. However, in our present work we have performed explicit computation of $T_c$ based on $T_1, ~T_2$ calculations for the whole range of $\rho_F$ from $0$ to $1$, as shown in figures (\ref{fig:fig4}) and (\ref{fig:fig5}). Figure \ref{fig:fig7} presents a plot of the ratio $T_c/B_0$ as a function of $\rho_F$. We note that the calculated value of the ratio for $\rho_F=1/2$ is in almost exact agreement with our earlier estimate\cite{zeroTarxiv, zeroTPRO}. The ratio is between $0.01$ and $0.10$ for most $\rho_F$ values in the range 0.3 to 1.0.

Figure \ref{fig:fig8} plots the ratio $T_c/B_c$ as a function of $\rho_F$, using $B_c = 2\psi_c^2$, the $T=T_c$ analogue of the $T=0$ Fermi band width $B_0$. It is interesting to compare this with experimentally determined  values of the ratio $T_c$/$T_F$, of the superfluid or superconducting transition temperature $T_c$ to the Fermi temperature $T_F$. This ratio also scales approximately with the ratio $\Delta$/$E_F$, the pairing strength of single-band superconductors, where $\Delta$ is the superconducting gap and $E_F$ is the Fermi energy. In particular, it has been recognized\cite{Uemura} that the unconventional superconductors show a relatively large value of $T_c$/$T_F \sim 0.02-0.2$, while the conventional superconductors show a small value of  $T_c$/$T_F$ $\sim$ $10^{-4}$ to $10^{-5}$. For example, elemental metal tin (Sn)\cite{Ashcroft} has a  $T_c$/$T_F$ $\sim$  3$\times$$10^{-5}$, while the high-$T_c$-cuprates\cite{Yamamoto, Brookes, Xie} exhibit a $T_c$/$T_F$ $\sim$ 0.02-0.03. On the other hand, the iron based superconductors (estimated from experimental data\cite{Lubashevsky, Okazaki, Kasahara, Rinott, MonoFeSe}) and the newly discovered hydride superconductors (as estimated from experimental $T_c$ values\cite{Drozdov, Drozdov2, Somayazulu} and band structure calculations\cite{Bianconi, Jarlborg, Liu}) show $T_c$/$T_F$ $\sim$ 0.1 - 0.2. Surprisingly, the corresponding value for a Fermi-Bose mixture\cite{Ferrier} estimated from experimental data is $T_c$/$T_F$ $\simeq  0.19$, although the $T_c$$\sim$ 200 nK. Even for a purely ultracold Fermi atomic system\cite{Chin}, $T_c$/$T_F$ $\simeq 0.2$. Thus, the high-$T_c$ cuprates, iron-based superconductors, hydrides and ultracold atomic systems are clearly classified as unconventional superconductors/superfluids \cite{Millev90}. 

A very recent study on an iron-based superconductor has reported evidence for the first solid-state BEC superconductor, although the observed value of the ratio $\Delta$/$E_F$ ($\simeq 0.12$, corresponding to a $T_c$/$T_F$ $\simeq  0.025$) is reduced across the BCS-BEC crossover, and the authors interpret it as most probably arising from interband coupling effects\cite{Hashimoto}.

The $T_c$ values of these unconventional materials span nine orders of magnitude: while the ultracold FBMs have $T_c \simeq 200$nK, the hydrides have a $T_c \simeq 250$K. But the important point is that the ratio calculated in our model is remarkably close to these values for several classes of unconventional superfluids and superconductors. In particular, we obtain $T_c/B_0 \sim 0.01-0.1 ~(T_c/B_c \sim 0.01-0.25)$; the value of this ratio is a very robust feature of our model, and compares well with the range $0.03-0.22$ for most unconventional materials (see the Table 1 of our previous paper for details\cite{zeroTPRO}).

An important question is that of a potential experimental realization of the composite hopping model. Consider a lattice with $N$ sites and $M$ electrons ($2N>M>N$) with nearest-neighbor hopping and on-site repulsion. In this case, by the Pigeonhole principle, at least one site must have more than one electron. Taking into account the repulsion between electrons, in the minimum energy configuration we have $M-N$ local pairs (that can be considered as hardcore bosons) and $2N-M$ electrons each occupiing a site. When one of the electrons of a pair hops to a neighboring site with one electron, we have a realization of composite hopping (see our earlier work\cite{zeroTarxiv, zeroTPRO} for a more detailed explanation). This situation is perhaps also well described by a one-band fermion Hubbard model above half filling, and the composite hopping model might offer a good effective description. Our model approximates the electrons by spinless fermions, but in subsequent work we plan to treat the spin half case. In this case, the composite hopping strength $t$ is just the single-electron hopping strength. For a narrow band Hubbard model, we can take $t \sim 0.1-0.2$eV, and if we further use $T_c \simeq 0.06 zt$ (corresponding to its peak value in Fig.\ref{fig:fig4} at the van Hove point $\rho_F=1/2$), we obtain a value of $T_c \simeq 250-500$K. 

If room-temperature superconductivity is theoretically possible within a composite-hopping framework, then it provides a strong reason to look for practical examples of it. In the pigeonhole context discussed above, external pressure might help overcome the repulsion between electrons and keep them paired. It is interesting to explore if high-pressure room-temperature superconductors like sulphides\cite{dias2020} are indeed solid-state realizations of the composite-hopping model, and are superconducting because of such a `pigeonhole' pairing mechanism.

A second possibility might be realization of composite hopping in the context of ultracold atoms in an optical lattice. Quantum simulation is an exciting area of research\cite{lewenstein2012}, and given the many unusual physical properties displayed by our model such as negative compressibility (discussed in our earlier work\cite{zeroTarxiv, zeroTPRO}), negative entropy\cite{cerf,delrio2011,Chatzi2020}, zeroth-order transition\cite{Maslov2004, Gunasekaran, Altamirano01, Zou, Hennigar, Altamirano02, Amin, kundu2020} and the remarkable agreement of the calculated ratio of  $T_c$/$T_F$  with a wide range of unconventional superfluids and superconductors \cite{zeroTPRO}, it would be interesting to explore the physics of composite hopping in this perspective.

In our model, there is a clear distinction between the boson-dominated ($\rho_F \le 1/2$) and the fermion-dominated ($\rho_F > 1/2$) regimes: in the former, the temperature-driven transition is first-order and the entropy remains positive throughout, while in the latter regime the transition is zeroth order and the entropy is negative in the ordered phase close to $T_c$. In the latter regime, the zero-temperature bulk modulus becomes negative over a range of $\rho_F$\cite{zeroTPRO}.

An important question whether the zeroth order temperature-driven transition for $\rho_F>1/2$ is a consequence of mean-field approximation (\ref{eq:mfa}), or an inherent property of the composite-hopping model. This question assumes importance given that the model offers an irreducible nontrivial description of an FBM perhaps not explored before, and the circumstance that zeroth-order transitions are relatively rare in the solid state\cite{Maslov2004, kundu2020}. It would therefore be of interest to study the model using other numerical techniques of quantum many-body theory like Quantum Monte-Carlo calculations (QMC)\cite{bhmqmc}.

\section{Conclusion}
\label{conclude}

In this work, we have extended the zero-temperature study of the composite hopping model of FBMs to explore finite-temperature thermodynamics. We computed the $\rho_F-T$ phase diagram and found that the temperature-driven metallic superfluid phase to insulating normal phase transition is discontinuous: for $\rho \le 1/2$, it is first order, while for $\rho>1/2$ it is zeroth order. We calculated the the temperature dependent superfluid amplitude $\psi$, the fermion hopping amplitude $\phi$, the Fermi chemical potential, and free energy within a mean-field approximation.  We also computed the entropy and found that it becomes negative for a certain range of temperatures below $T_c$ when $\rho_F>1/2$. The ratios $T_c/B_0$ and $T_c/B_c$ where $B_0$ and $B_c$ are Fermi band widths at $T=0$ and $T=T_c$, respectively, could also be calculated.
The calculated ratios of $T_c$ to Fermi band widths match well with experimental values of $T_c$/$T_F$ (where $T_F$ is the Fermi temperature) of unconventional superfluids and superconductors, including Fermi-Bose mixtures, the high-$T_c$ cuprates and iron-based superconductors, that span a $T_c$ range of nine orders. The results indicate the important role of composite hopping in describing the superfluid properties of Fermi-Bose mixtures. 

\newpage

\section*{Acknowledgments}
A.C. thanks the Ministry of Science and Technology of the Republic of China, Taiwan, for financially supporting this research under Contract No. MOST 108-2112-M-213-001- MY3.

\end{document}